\documentstyle[12pt]{article}
\input{epsf}
\newcommand{\be}{\begin{eqnarray}}
\newcommand{\ee}{\end{eqnarray}}
\newcommand{\ba}{\begin{array}}
\newcommand{\ea}{\end{array}}

\begin{document}

\rightline{\textsl{\date{\today}}} \vspace{0.5cm}
\begin{center}
{\Large The neutron anomaly in  the $\gamma N\to\eta N$  cross section through the looking glass of the flavour SU(3) symmetry}\\
\vspace{0.35cm}
 T. Boiko$^1$,  V. Kuznetsov$^{2}$ and M.V. Polyakov$^{2,3}$\footnote{e-mail address: maxim.polyakov@tp2.rub.de}\\

\vspace{0.35cm}
$^1$ Bodwell High School,
955 Harbourside Drive,
Vancouver, BC,
Canada V7P 3S4 \\
$^2$ Petersburg Nuclear Physics Institute, Gatchina, 188300, St. Petersburg, Russia\\
$^3$Institute f\"ur Theoretische Physik II, Ruhr-Universit\"at Bochum,
D - 44780 Bochum, Germany

\end{center}

\begin{abstract}
\noindent
We study the implications of the flavour SU(3) symmetry for various interpretations of the neutron anomaly  in  the $\gamma N\to\eta N$
cross section. We show that the explanation of the neutron anomaly due to interference of known N(1535) and N(1650) resonances 
implies that N(1650) resonance should have a huge coupling to $\phi$-meson -- at least 5 times larger than the corresponding  $\rho^0$ coupling.
In terms of quark degrees of freedom this means that the well-known N(1650) resonance must be  a ``cryptoexotic pentaquark"--
its wave function should contain predominantly an $s\bar s$ component.

\noindent
It turns out that the ``conventional" interpretation of the neutron anomaly by the interference of known resonances 
metamorphoses into unconventional physics picture  of N(1650).
\end{abstract}

\section*{\normalsize \bf Introduction}

\begin{figure}[h]
\vspace*{0.6cm}
\centerline{\epsfverbosetrue\epsfxsize=7.5cm\epsfysize=5.cm\epsfbox{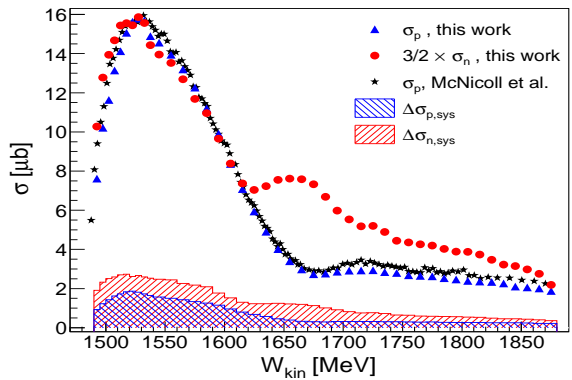}}
\caption{\small Figure from Ref.~\cite{wert}. Total cross sections as a function of the final-state invariant mass 
$m(\eta N)$: Blue triangles: proton data. Red circles: neutron data scaled by $3/2$. 
Black stars: free proton data from MAMI-C \cite{Mainz}.
Hatched areas: total systematic uncertainties of proton (blue) and neutron (red) data.}
    \label{fig:krusche} \vspace{0.3cm}
\end{figure}

The discovery of the neutron anomaly\footnote{ Existence of the narrow ($\Gamma\sim$10-40~MeV) peak in the $\gamma n\to \eta n$ cross section around 1680~MeV and its absence 
in the $\gamma p\to \eta p$ process} in  the $\gamma N\to\eta N$  cross section was reported in Ref.~\cite{gra0},  in this paper
  the GRAAL data on
the photon scattering off the deuteron were analysed.
Presently three other collaborations  
( LNS~\cite{kas},CBELSA/TAPS\cite{kru}, and A2~\cite{wert}) confirmed the neutron anomaly beyond any doubts.  
For an illustration of the neutron anomaly in $\gamma N\to\eta N$ we show  on Fig.~\ref{fig:krusche} the most recent results of the A2 collaboration 
\cite{wert}. 
Furthermore the neutron anomaly at the same invariant mass of $W\sim 1680$~MeV 
was also observed in the Compton scattering \cite{Compton}. 

In our view the observation of the neutron anomaly is the most striking discovery in the field of the nucleon resonances spectroscopy during the last decade. It is
important to figure out the physics nature of the phenomenon.
In the present paper we study the implications of the flavour SU(3) symmetry for various explanations of the neutron anomaly. \\


\section*{\normalsize \bf Flavour SU(3) decomposition of the $\gamma N\to \eta N$ amplitude}

In the SU(3) symmetry limit the amplitudes $A\left(\gamma p\to \eta p\right)$ and $A\left(\gamma n\to \eta n\right)$ can be decomposed
through the amplitudes corresponding to the irreducable representations of the SU(3) group in the s-channel. The photon and the nucleon belong
to the octet representation of the SU(3) group. Therefore the possible representations in the s-channel are those which appear in the 
product ${\bf 8}\times {\bf 8}={\bf 1}+{\bf 8}_{\rm F}+{\bf 8}_{\rm D}+{\bf 10}+{\bf \overline{ 10}}+{\bf 27}$. Obviously, the $\bf 1$ and $\bf 10$ representations do not
enter the decomposition of the $\gamma N\to \eta N$ amplitude. The SU(3) decomposition for the amplitudes has the following form:
\be
A\left(\gamma p\to \eta p\right)&=&\frac 13 A_{\rm D}^{(8)}+A_{\rm F}^{(8)}+A^{(27)},
\nonumber\\
A\left(\gamma n\to \eta n\right)&=&\frac 23 A_{\rm D}^{(8)}+A^{(\overline{ 10})}+\frac 12 A^{(27)}.
\label{decomposition}
\ee
One sees that the anti-decuplet amplitude $A^{(\overline{ 10})}$ do not enter the $\gamma p$ channel, whereas the antisymmetric octet amplitude
$A_{\rm F}^{(8)}$ do not enter the $\gamma n$ channel.

In order to describe the phenomenon of the neutron anomaly one needs that the amplitude $A\left(\gamma n\to \eta n\right)$ is very different (larger size and 
more rapid  energy dependence) from $A\left(\gamma p\to \eta p\right)$ on a narrow invariant energy interval (several tens of MeV) around $W\sim 1680$~MeV.
The decomposition (\ref{decomposition}) 
offers three possibilities to arrange such difference (ordered according to the {\it Prinzip der Denk\"okonomie}): 
\begin{itemize}
\item[(I)] the anti-decuplet amplitude $A^{(\overline{ 10})}$ has large size and rapid energy dependence on a narrow energy interval
around 1680~MeV,
\item[(II)] there is a conspiracy and a fine tuning among the SU(3) amplitudes $A_{\rm F}^{(8)},A_{\rm D}^{(8)}$ and $A^{(27)}$ on that narrow energy interval,
\item[(III)] an extraordinarily strong violation of the SU(3) symmetry on that narrow energy interval.
\end{itemize}
We emphasise that the option (II) can explain the neutron anomaly  only in the $\eta$-photoproduction. In other channels, e.g. the Compton scattering \cite{Compton},
the  assumed conspiracy and fine tuning are destroyed due to different from (\ref{decomposition})   SU(3) decomposition of the Compton amplitude.
The anti-decuplet amplitude $A^{(\overline{ 10})}$ enters only the $\gamma n$ channel independently of the final state. Therefore
the option (I)  predicts the neutron anomaly for the Compton scattering as well.

Usually the approximate flavour SU(3) symmetry works pretty well. As a rule its predictions are satisfied with an accuracy of about 30\% or better, see e.g. 
a review \cite{SU3}. A very large violation of the SU(3) symmetry would be a serious challenge to our common wisdom about hadron dynamics.

It is likely that a possible realisation of the option (III) is provided by  Ref.~\cite{dor}. In this paper the neutron anomaly was explained by the threshold
 effect due to $K\Lambda$ and $K\Sigma$ intermediate states. It was argued in Ref.~\cite{dor} that the intermediate
$K^+ \Sigma^-$ state in the $\gamma n $ channel produces the cusp effect at $W\sim 1685$~MeV which can explain the peak in that channel. In order to suppress
the corresponding peak in the $\gamma p$ channel the authors of Ref.~\cite{dor} fitted their model parameter in such a way that the cusp due to $K^+\Lambda$ intermediate state
cancels the cusp effect due to $K^+\Sigma^0$ state (a kind of fine tuning). Such fine cancellation may require a  large violation of the SU(3) symmetry. We shall consider 
this case in details elsewhere \cite{ahlborn}. In any case, the explanation of the neutron neutron anomaly of Ref.~\cite{dor} is not universal, {\it i.e.} it works 
only for $\eta$-photoproduction and fails for Compton scattering, the same as for the option (II).

In the following sections we analyse the physics realisations of the first two possibilities discussed above.\\

\section*{\normalsize \bf (I)  Dominance of the anti-decuplet amplitude }

The simplest physics realisation of the option (I) is an existence of a narrow anti-decuplet of baryons. The existence of such narrow exotic baryon multiplet
was predicted in Ref.~\cite{dia}. 
Main properties of N$^*$ from the anti-deculpet which were predicted theoretically in years 1997-2004 (before the discovery of the neutron anomaly) are the following:
\vspace{-0.2cm}
\begin{itemize}
\item quantum numbers are $P_{11}$ ($J^P=\frac 12^+$, isospin=$\frac 12$) \cite{dia},
\vspace{-0.2cm}
\item narrow width of $\Gamma\le 40$~MeV \cite{dia,arndt,michal},
\vspace{-0.2cm}
\item mass of $M\sim 1650-1720$~MeV \cite{arndt,michal,dia1},
\vspace{-0.2cm}
\item strong suppression of  the proton photocoupling relative to the neutron one \cite{max} ,
\vspace{-0.2cm}
\item the $\pi N$ coupling  is suppressed, N$^*$  prefers to decay into $\eta N$, $K\Lambda$ and $\pi \Delta$ \cite{dia,arndt,michal}.
\end{itemize}
The nucleon resonance with such properties can explain\footnote{Actually it was prediction of the phenomenon.}  concisely the neutron anomaly in $\eta$-photoproduction, in the Compton scattering,
 in the $\gamma N\to K \Lambda$ process, etc.

Detailed account for predictions and evidences for narrow anti-decuplet nucleon 
was presented at length previously in the literature (see e.g. \cite{acta,micha}). Not to dwell on this again, we just list
the extracted properties of the putative anti-decuplet nucleon resonance (and relevant references) in  Table~\ref{tab:alapdg}.  \\

{\small
\begin{table}[tb]
  \begin{center}
    \begin{tabular}{lccc}  
    observable   &     extracted value &      refs. (neutron data) & refs. (proton data)  \\
    \hline
      mass (MeV)      &$1680\pm 15$  & \cite{gra0,kas,kru,wert,Compton}\cite{arndt}$^{\star)}$& \cite{acta,jetp,KPT,BG}  \cite{arndt}$^{\star)}$\\
     $\Gamma_{\rm tot}$ (MeV)  & $\le 40$     &    \cite{gra0,kas,kru,wert,Compton}\cite{arndt}$^{\star)}$& \cite{acta,jetp,KPT,BG} \cite{arndt}$^{\star)}$ \\
      $\Gamma_{\pi N}$ (MeV) &    $\le 0.5$ &      \cite{arndt}$^{\star)}$&  \cite{arndt}$^{\star)}$ \\
      $\sqrt{{\rm Br}_{\eta N}} A_{1/2}^n \ (10^{-3}\ {\rm GeV}^{-1/2})$   & 12-18     &      \cite{az,wert} & \\
$\sqrt{{\rm Br}_{\eta N}} A_{1/2}^p \ (10^{-3}\ {\rm GeV}^{-1/2})$         & 1-3         &                                  &   \cite{acta,jetp,KPT,BG}  \\
    \end{tabular}
    \caption{ {\small Our estimate of properties of the putative narrow N$^*$ extracted from the data under the assumption that N$^*$ exists.  
    $^{\star)}$In Ref.~\cite{arndt} the elastic $\pi N$ scattering data were analyzed
    and the tolerance limits for N$^*$ parameters were obtained. The preferable  quantum numbers in this analysis are $P_{11}$.}}
    \label{tab:alapdg}
  \end{center}
\end{table} }
%
\vspace{-0.5cm}

\section*{\normalsize \bf (II) Conspiracy and fine tuning among non-exotic SU(3) amplitudes}

A physics realisation of the option (II) was suggested in Refs.~\cite{ani,ani1,ani2} by the Bonn-Gatchina group (BnGa). In these papers the neutron anomaly was explained
by the interference effect of well-known wide S$_{11}$ resonances N(1535) and N(1650). In order to arrange a narrow structure in the neutron channel the photocouplings
of these two resonances should be fine tuned. In particular, the proton and neutron photocouplings of N(1650) must have the same sign. To describe the most recent
and the most precise data of the A2 collaboration  on the neutron anomaly \cite{wert} BnGa obtained the following ratio of the proton to neutron
photocouplings \cite{ani2}:
\be
R_{pn}\equiv \frac{A^p_{1/2}(1650)}{A^n_{1/2}(1650)}=1.74\pm0.66 \ \ \ {\rm [BnGa\ value]}.
\label{BGratio}
\ee 
Employing the flavour SU(3) symmetry one can express the ratio of the $F_V$ and $D_V$ octet vector couplings in terms of the ratio $R_{pn}$:

\be
\frac{F_V}{D_V}=-\frac 13 \left(2 R_{pn} +1\right)=-1.50\pm0.44 \ \ \ {\rm [BnGa\ value]}.
\label{FkDR}
\ee
The resulting from the analysis \cite{ani2} $F_V$ to $D_V$ ratio is negative and larger than 1 in the absolute value. To our best knowledge such
values of $F_V/D_V$ have been never obtained in any model of baryon resonances (variants of quark model, MIT bag model, soliton models, etc). 
Let us see what are physics implications of such unusual values of the $F_V/D_V$ ratio.

The flavour SU(3) symmetry allows to express various flavour combinations of the vector current couplings in terms of $F_V/D_V$-ratio (and hence in terms
of $R_{pn}$ (\ref{BGratio}) owing Eq.~(\ref{FkDR})). One can easily derive the following relations for various vector couplings of N(1650) (valid also for 
any octet nucleon resonance $N'$):
\be
\label{rom}
R_\omega&\equiv&\frac{g_{\omega N N'}}{g_{\rho^0 N N'}}=\frac{R_{pn}+1}{R_{pn}-1}+\sqrt{\frac 2 3}\  r_0, \\
\label{rfi}
R_\phi&\equiv&\frac{g_{\phi N N'}}{g_{\rho^0 N N'}}=-\sqrt 2\  \frac{R_{pn}+1}{R_{pn}-1}+\sqrt{\frac 1 3}\  r_0.
\ee  
Here $r_0$ is the ratio of the flavour singlet ($(\bar u\gamma_\mu u+\bar d\gamma_\mu d+\bar s\gamma_\mu s)/\sqrt 3$) vector current
coupling to the isovector ($(\bar u\gamma_\mu u-\bar d\gamma_\mu d)/\sqrt 2$) that. The value of $r_0$ is not fixed by the SU(3) symmetry, however
with help of Eqs. (\ref{rom},\ref{rfi}) we can express $\phi$-meson coupling $R_\phi$ in terms of the $\omega$-meson coupling $R_\omega$
and the proton to neutron ratio of the photocoupling $R_{pn}$:
\be
\label{romrfi}
R_\phi=\frac{1}{\sqrt 2} \ \left(R_\omega-3\ \frac{R_{pn}+1}{R_{pn}-1}\right).
\ee 
Additionally, from Eqs. (\ref{rom},\ref{rfi}) one can easily derive the following inequality (kind of Cauchy-Bunyakovsky-Schwarz inequality):

\be
\label{wvarc}
R_\omega^2+R_\phi^2 \geq 3\ \left( \frac{R_{pn}+1}{R_{pn}-1} \right)^2.
\ee
If we take the BnGa value (\ref{BGratio}) for $R_{pn}$ we obtain from (\ref{wvarc}):
\be
R_\omega^2+R_\phi^2 \geq 27.
\ee
One sees that in the scenario of Refs.~\cite{ani,ani1,ani2} the $\omega$- and $\phi$-meson couplings of N(1650)
can not be small simultaneously\footnote{Note that if we take N(1650) photocouplings from the SAID analysis \cite{said}, than
$F_V/D_V =0.6\pm0.6$ (range of values typical for all models of baryon resonances) and $R_\omega^2+R_\phi^2 \geq 0.2$ (small values of $\omega$
and $\phi$ couplings are not excluded).}.

Experimentally \cite{PDG} the decay ${\rm N}(1650)\to \rho N$ is seen and sizable, however the decay ${\rm N}(1650)\to \omega N$ is not seen.
If we conservatively assume that the yield of $\omega$ mesons does not exceed factor of four relative to the yield of $\rho$ mesons, {\it i.e}
$R_\omega^2\leq 4$
then from Eq.~(\ref{romrfi}) with BnGa value  for the $p/n$ ratio of N(1650) photocouplings (\ref{BGratio}) we obtain:
\be
\left| R_\phi\right| \geq 6.
\ee  
We see that the explanation of the neutron anomaly by the interference of known N(1535) and N(1650) resonances advocated in \cite{ani,ani1,ani2}
suggests that the $\phi$-meson (almost pure $s\bar s$ state) coupling to N(1650)$\to$N transition should be huge. In terms of quark degrees of freedom
it means that N(1650) has a large admixture of $s\bar s$ component, {\it i.e.} in the scenario of Refs.~\cite{ani,ani1,ani2} N(1650) is dominantly  ``cryptoexotic
pentaquark" .  It turns out that the ``conventional" interpretation of the neutron anomaly by the interference of known resonances \cite{ani,ani1,ani2} 
metamorphoses into unconventional physics picture  of N(1650).

\section*{\normalsize \bf Conclusions}

In summary, we analysed the implication of the flavour SU(3) symmetry for explanations of the neutron anomaly in the $\gamma N\to \eta N$ cross section.
The SU(3) symmetry suggests two class of scenarios: (I) dominance of the anti-decuplet channel at narrow energy interval (II) fine tuning of parameters of
known wide resonances to arrange very specific interference pattern, see Refs.~ \cite{ani,ani1,ani2}. 

Both scenarios need exotic nucleon resonances -- this can be either (I) a narrow anti-decuplet of baryons, or (II) well know N(1650) resonance with very 
large
$s\bar s$ component, {\it i.e.} the well-known N(1650) resonance must be  a ``cryptoexotic pentaquark".
 
 We stressed that the option (II) (in contrast to the first scenario) can explain the neutron anomaly only in the $\gamma N\to \eta N$ process and it fails to explain the neutron anomaly in the Compton scattering. 
It seems that the simplest, universal  (for both $\eta$-photoproduction and the Compton 
scattering) and concise way to explain the neutron anomaly is the existence of a narrow anti-decuplet of baryons.


\section*{\normalsize \bf Acknowledgements}
We are grateful to Y.  Azimov, M.~D\"oring, A.~Gasparyan, J.~Gegelia, B.~Krusche, A.~Sarantsev,  and I. Strakovsky for stimulating discussions. 
The work of TB and VK was partially supported by DFG (SFB/TR16 grant), and by the administration of High Energy Physics Department of PNPI.


\end{document}